%% file: mydocument.tex
\newcommand*{\ATLASLATEXPATH}{latex/}
\documentclass[cernpreprint,texlive=2011,txfonts,UKenglish]{\ATLASLATEXPATH atlasdoc}
 \pdfoutput=1

\usepackage{\ATLASLATEXPATH atlaspackage }

\usepackage{\ATLASLATEXPATH atlascontribute}

\usepackage{\ATLASLATEXPATH atlasphysics}

\graphicspath{{logos/}{figures/}}

\usepackage{mydocument-defs}

\hypersetup{pdftitle={ATLAS draft},pdfauthor={The ATLAS Collaboration}}
\AtlasTitle{Search for a new resonance decaying to  a $W$ or $Z $ boson and a Higgs boson in the $\ell \ell/ \ell \nu/ \nu \nu + b \bar{b}$ final states  with the ATLAS Detector}
\AtlasAbstract{                                                                                                                                                                
A search  for a  new resonance decaying to  a $W$ or $Z$ boson and a Higgs boson  in the $\ell \ell/ \ell \nu/ \nu \nu + b \bar{b}$ final states is performed using  20.3 fb$^{-1}$ of $pp$ collision data recorded at $\sqrt{s}=$~8~TeV with the ATLAS detector at the Large Hadron Collider. The search is conducted by examining the $WH/ZH$ invariant mass distribution for a localized excess.
No significant deviation from the Standard Model background prediction is observed. The results are interpreted 
in terms of constraints on the Minimal Walking Technicolor model and on a simplified approach based on a phenomenological Lagrangian  of Heavy Vector Triplets.
}

\AtlasJournal{Eur. Phys. J. C.}
\PreprintIdNumber{CERN-PH-EP-2015-052}

\begin{document}

\maketitle

\section{Introduction}
\label{sec:intro}
\input{introduction}

\section{The ATLAS detector}
\label{sec:det}
\input{detector}

\section{Data and Monte Carlo samples}
\label{sec:samples}
\input{samples}

\section{Object reconstruction}
\label{sec:objects}
\input{objects}

\section{Event selection and reconstruction}
\label{sec:selection}
\input{selection}

\section{Background estimation}
\label{sec:sigbkg}
\input{sigbkg}

\section{Systematic uncertainties}
\label{sec:systs}
\input{systs}

\section{Results and limit extraction}
\label{sec:results}
\input{results}

\FloatBarrier

\section{Summary}
\label{sec:summary}
\input{summary}

\section*{Acknowledgements}

\input{acknowledgements/Acknowledgements}

\bibliographystyle{bibtex/bst/atlasBibStyleWoTitle}
\bibliography{mydocument}

\newpage 
\input{atlas_authlist}
\end{document}

%% file: introduction.tex
Although the Higgs boson discovery by the \mbox{ATLAS}~\cite{:2012gk} and CMS~\cite{:2012gu} collaborations imposes strong constraints on theories beyond the Standard Model (SM), 
the extreme fine tuning in quantum corrections required to have a light fundamental Higgs boson~\cite{Agashe:2014kda,Wilson:1970ag} suggests that the SM may be incomplete, and 
not valid beyond a scale of  a few TeV. Various dynamical electroweak symmetry breaking scenarios which attempt to solve this naturalness problem, such as Minimal Walking Technicolor~\cite{Sannino:2004qp,Foadi:2007ue,Belyaev:2008yj,Cacciapaglia:2014uja}, Little Higgs~\cite{Schmaltz:2005ky}, or composite Higgs models~\cite{Dugan:1984hq,Agashe:2004rs}, predict the existence of new resonances decaying to a vector boson plus a Higgs boson.  

Using the full dataset collected by the ATLAS detector at 8 TeV centre-of-mass energy at the Large Hadron Collider, a search is performed for a heavy resonance decaying to $VH$, where $V$ is a $W$ or $Z$ boson and $H$ is the SM Higgs boson. This analysis looks for the leptonic decay of the $W$ or $Z$ boson and the Higgs decay into a $b$-quark pair. Therefore the selected final states are: zero charged leptons   targeting $Z(\rightarrow \nu\nu)b\bar{b}$ decays, one charged lepton  $W(\rightarrow \ell\nu)b\bar{b}$, and two oppositely charged leptons $Z(\rightarrow \ell\ell)b\bar{b}$ where $\ell =e, \mu$.  The search is performed by examining the distribution of the reconstructed $VH$ mass  ($m_{VH}$)  for a localized excess. The signal strength and the background normalization are determined from a likelihood fit  to the data distribution in the three  channels studied.

As a benchmark, the Minimal Walking Technicolor model (MWT) is used, a model with  strongly coupled dynamics.
This model predicts two triplets of resonances, $R_1^{\pm,0}$ and $R_2^{\pm,0}$, one of which is a vector and the other an axial-vector, that couple to vector bosons with strength $\tilde g$ and to fermions with $g/\tilde g$, where $g$ is the weak SU(2) coupling constant. The  bare axial-vector mass $m_A$  determines the masses of $R_1$ and $R_2$, with the lower mass resonance $R_1$ having a mass close to $m_A$.  Recent  lattice simulations in this model~\cite{Fodor:2012ty,Fodor:2014pqa,Hietanen:2014xca} predict masses close to 2 TeV.
 The decay channels $R_{1,2}^{\pm} \rightarrow WH$ and $R_{1,2}^{0} \rightarrow ZH$, lead to  $Wb\bar{b}$ and $Zb\bar{b}$ final states.

A simplified approach based on a phenomenological Lagrangian~\cite{Pappadopulo:2014qza} that incorporates Heavy Vector Triplets (HVT), which allows  the interpretation of the results in a model-independent way, is also used. Here, the new heavy vector bosons, $V'^{\pm,0}$, couple to the Higgs and SM gauge bosons via a combination of parameters $g_Vc_H$ and to the fermions via the combination $(g^2/g_V)~c_F$. 
The parameter $g_V$ represents the strength of the new vector boson interaction, while $c_H$ and $c_F$, which represent the couplings to the Higgs and the fermions respectively,  are expected to be of  order  unity in most models. 
Two benchmark models~\cite{Pappadopulo:2014qza} are used here. 
In the first model, referred to  as \textit{model A}, the branching fractions to fermions and gauge bosons are comparable, as in some extensions of the
SM gauge group~\cite{Barger:1980ix}.  For \textit{model B}, fermionic couplings are suppressed, as for example in a composite  Higgs
model~\cite{Contino:2011np}.
%

The three final states presented in this Letter have been extensively studied for non-resonant production  in ATLAS~\cite{AtlasVHbb}. Moreover, a search for a pseudoscalar resonance in the $\ell \ell b \bar{b}$ and $\nu \nu b \bar{b}$ channels has already been published by ATLAS, setting limits on two-Higgs-doublet models~\cite{Aad:2015wra}.
Other searches for particles occurring in MWT and HVT models have been conducted by the ATLAS~\cite{Aad:2014pha,Aad:2014cka} and CMS~\cite{Khachatryan:2015ywa} collaborations.

%% file: detector.tex
 The ATLAS detector~\cite{1748-0221-3-08-S08003} is a general-purpose particle detector used to investigate a broad range of physics processes. It includes inner tracking devices surrounded by a superconducting solenoid, electromagnetic and hadronic calorimeters and a muon spectrometer. The inner detector (ID) provides precision tracking of charged particles with 
 pseudorapidity\footnote{{ATLAS} uses a right-handed coordinate
system with its origin at the nominal interaction point (IP) in the centre of the detector and
the $z$-axis along the beam pipe. The $x$-axis points from the IP to the centre of the LHC ring,
and the $y$-axis points upward. Cylindrical coordinates $(r,\phi)$ are used in the transverse
plane, $\phi$ being the azimuthal angle around the beam pipe. The pseudorapidity is defined in
terms of the polar angle $\theta$ as $\eta=-\ln\tan(\theta/2)$.}
 ~$|\eta| < 2.5$.
The calorimeter system covers the pseudorapidity range $|\eta| < 4.9$. It is composed of sampling calorimeters with either liquid argon (LAr) or scintillator tiles as the active medium. 
The muon spectrometer consists of three large superconducting toroids and a system of trigger chambers and precision tracking chambers that provide triggering and tracking capabilities in the ranges  of $|\eta| <$ 2.4 and  $|\eta| < $2.7 respectively.

The ATLAS detector has a three-level trigger system to select events for offline analysis.

%% file: samples.tex
This analysis is based on $\sqrt{s}=8\TeV$ $pp$ collision data corresponding to 20.3~$\pm$~0.6~\ifb ~\cite{ATLAS_Luminosity2012}. 
The data used in the $\ell \nu b\bar{b}$ final state were collected using single-electron and single-muon triggers with transverse momentum ($\pt$) thresholds  from 24 to 60~GeV.
The data used in the $\ell\ell b\bar{b}$ final state were collected using a combination of single-electron, single-muon, dielectron ($ee$) and dimuon  ($\mu\mu$) triggers.
The $\pt$ thresholds for the $ee$ and $\mu\mu$ triggers vary  from 12 to 13~GeV.
The data used in the $\nu\nu b\bar{b}$ final state were collected using a trigger that requires a missing transverse momentum ($\vec{\met}$)  with magnitude $\ETMISS$ greater than 80~GeV.	

Simulated Monte Carlo (MC) samples for the MWT benchmark model use the implementation~\cite{MWT-tools} in {\sc Madgraph5}~\cite{Alwall:2011uj}, with the Higgs boson mass set to 126 GeV. 
The parameter $\tilde g$ is set to 2 for  signal generation. Constraints on other values of this parameter can be set using the same samples since  the kinematic distributions do not depend on $\tilde g$. The parameter $S$, which is an approximate value~\cite{Foadi:2012ga} of the Peskin--Takeuchi $S$ parameter~\cite{Peskin:1991sw} which measures potential new contributions to electroweak radiative corrections, is set to 0.3, in accordance with the recommendations in Ref.~\cite{Andersen:2011nk}.

Signal samples for the HVT model are also generated with {\sc Madgraph5}. The parameter $c_F$  is assumed to be the same for quarks and leptons including third-generation fermions. Other parameters involving more than one heavy vector boson, $g_Vc_{VVV}$,  $g_V^2 c_{VVHH}$ and $c_{VVW}$, have negligible effect on the overall cross sections for the processes of interest here.
For all signal events, parton showering and hadronization  is performed with {\sc Pythia8} \cite{Pythia,Pythia8} and the {\sc CTEQ6L1}~\cite{Pumplin:2002vw} parton distribution functions  (PDFs) are used. Benchmark signal samples are generated for a range of resonance masses from $300$ to $2000\,\GeV$ in steps of 100~GeV.

MC samples are used to model the shape and normalization of most SM background processes, although some are later adjusted using data-based corrections extracted from control samples.	 The production of $W$ and $Z$ bosons in association with jets is simulated with {\sc Sherpa} 1.4.1~\cite{SHERPA} using the {\sc CT10} PDFs~\cite{PhysRevD.82.074024}.
Top quark pair  production is simulated using {\sc Powheg}~\cite{Powheg1,Powheg2} with the {\sc Powheg-BOX} program~\cite{Powheg3} interfaced to {\sc Pythia6}, using the {\sc CTEQ6L1} PDFs.
In this analysis, the final normalizations of these dominant backgrounds are constrained by the data, but theoretical cross sections are used to optimize the selection.
The cross sections are calculated at NNLO accuracy  for $W/Z$+jets~\cite{Melnikov:2006kv} and at NNLO+NNLL accuracy for $t\bar{t}$~\cite{Czakon:2013goa}.
Single top quark production is simulated with {\sc Powheg} and {\sc AcerMC}~\cite{AcerMC} interfaced to {\sc Pythia6}, using the {\sc CTEQ6L1} PDFs, and the cross sections are taken from Ref.~\cite{Kidonakis:2011wy}.
Diboson production ($WW$,$WZ$,$ZZ$) is simulated using {\sc Powheg} interfaced to {\sc Pythia8}, using the {\sc CT10} PDFs, and the cross sections
are obtained at NLO from {\sc mcfm}~\cite{Campbell:2010ff}.
Finally, SM Higgs boson production in association with a $W/Z$ boson is simulated using {\sc Pythia8} with the {\sc CTEQ6L1} PDFs, and considered as a background in this search. It is scaled to the SM cross section~\cite{AtlasVHbb}. 
	
All MC simulated samples include the effect of multiple $pp$ interactions in the same and neighbouring bunch crossings (pile-up) by overlaying simulated minimum-bias events on each generated signal or background event. The number of overlaid events is such that the distribution of the number of  interactions per $pp$ bunch crossing in the simulation matches that observed in the data, with on average $21$ interactions per bunch crossing. The generated samples are processed through the {\sc Geant4}-based ATLAS detector simulation~\cite{AGO-0301,ATLASSIM} or a fast simulation using a parameterization of the performance of the calorimetry and {\sc Geant4} for the other parts of the detector~\cite{FastCaloSim}.
Simulated events are reconstructed with the standard ATLAS reconstruction software used for collision data.

%% file: objects.tex
The physics objects used in this analysis are electrons, muons, jets and missing transverse momentum.

Electrons are identified for $|\eta| < 2.47$ and $\pt > 7$~GeV from
energy clusters in the electromagnetic calorimeter that are matched to
tracks in the inner detector~\cite{egammapaper}.  Quality requirements
based on the calorimeter cluster and track are applied to
reduce contamination from jets.

Muons are reconstructed in the muon spectrometer in the
range $|\eta|< 2.7$ and $\pt > 4$~GeV~\cite{Aad:2014rra}. 
For $|\eta|< 2.5$ the muon spectrometer track must be matched with 
a track in the inner detector and information from both is used to reconstruct
the momentum. Muons considered for this analysis must have $\pt > 7$~GeV.

Lepton candidates are required to be isolated to  reduce the multijet background.
The scalar sum of the transverse momenta of tracks with $\pt>1\,\GeV$ within a cone of $\Delta R=\sqrt{(\Delta\eta)^2+(\Delta\phi)^2}=0.2$ around the lepton track (tracking isolation) is required to be less than 10\% of the lepton $\pT$. 

Jets are reconstructed using the anti-$k_t$ algorithm~\cite{AntiKt} with radius parameter \mbox{$R = 0.4$}.
The jet transverse momentum is corrected for energy losses in passive
material, for  the non-compensating response of the calorimeter, and
for any additional energy due to multiple $pp$
interactions~\cite{JES}. Jets are required to have $\pT>30\,\GeV$ and
$|\eta|<4.5$. To reject low-$\pt$ jets from pile-up, for jets with
$\pT<50\,\GeV$ and $|\eta|<2.5$, the scalar sum of the $\pT$ of
associated tracks, originating from the reconstructed primary vertex,
is required to be at least 50\% of the scalar sum of the $\pT$ of all associated tracks.
To avoid double-counting of leptons and  jets, an overlap removal procedure is applied~\cite{AtlasVHbb}.

In the pseudorapidity range $|\eta|<2.5$, jets originating from
$b$-quarks are identified using a  multi-variate $b$-tagging
algorithm~\cite{ATLAS-CONF-2014-004}. This has an efficiency of 70\%
and a misidentification rate of less than 1\% for selecting jets
initiated by light quarks or gluons and of about 20\% for jets
initiated by $c$-quarks, as determined from $\ttbar$ MC events.
%
%

The missing transverse momentum is calculated as the negative of the vectorial sum of the calorimeter-based transverse momenta of all electrons,  jets, and calibrated calorimeter clusters within $|\eta| < 4.9$ that are not associated with any other objects~\cite{atlas_met}, as well as muon momenta. In addition, a track-based missing transverse momentum ($\vec{\mpt}$, with magnitude $\mpt$ ) is used, calculated as the negative vectorial sum of the track-based transverse momenta of objects
with $|\eta| < 2.4$ associated with the primary vertex.
%
%
%

%% file: selection.tex
Events are categorized into the $\nu\nu b\bar{b}$, $\ell\nu b\bar{b}$
or $\ell\ell b\bar{b}$ channels if they have zero, one or two reconstructed
charged leptons respectively. All categories require at least two jets in the pseudorapidity range $|\eta|< 2.5$ (central jets). 
The channels are further subdivided into
categories of events containing one or two $b$-tagged jets; events with
zero or $\geq 3$ $b$-tagged  jets are rejected.  The Higgs boson candidate (and its mass $m_{b\bar{b}}$)
 is reconstructed from the two $b$-tagged jets or, for 1-$b$-tag events, the $b$-tagged jet and the highest-$\pt$ remaining central
jet. In order to suppress $W/Z$+jets background, at least one of the jets must
have $\pt>45$~GeV and the invariant mass of the dijet pair must be  in the range $105<m_{b\bar{b}}<145$~GeV, consistent with the 
 Higgs mass.  In order to reduce the 
$t\bar{t}$ background in the $\nu\nu b\bar{b}$ and $\ell\nu b\bar{b}$
channels, events are rejected if they contain four or more jets.
To improve the resolution of the $VH$ mass a constraint to the Higgs boson mass is applied by scaling the Higgs boson candidate jet momenta  by $m_{H}/m_{b\bar{b}}$ ($m_{H}=125~\text{GeV}$). 
Further channel-specific cuts are applied as outlined below.

\subsection{ $\nu\nu b\bar{b}$ channel}
Events are selected with $\met>120$~GeV and $\mpt>30$~GeV.
A requirement is made on $H_\mathrm{T}$, defined as the scalar sum of the $\pt$ of 
all jets, in
 order to keep a high trigger efficiency: 
 $H_\mathrm{T}>120$~GeV ($>150$~GeV) for events with 
two (three) jets. Selections are also applied on the angle between the jets used for reconstructing the Higgs candidate, $\Delta R_{b\bar{b}}$, to suppress the $W/Z+$jets 
background~\cite{AtlasVHbb}: for $120 < \met < 160$~GeV, $0.7< \Delta R_{b\bar{b}} <1.8$; for $160 < \met < 200$~GeV, $\Delta R_{b\bar{b}}< 1.8$; for 
$\met >200$~GeV, $\Delta R_{b\bar{b}}<1.4$. Events containing an electron or muon passing the selection cuts described in Sec.~\ref{sec:objects} are removed.

In events with real \met\, the directions of \metvec\ and \vec{\mpt}
are expected to be similar.
In events with fake \met\ arising from a jet energy fluctuation, the direction of
\vec{\met} should be close to the direction of the poorly measured jet.
Therefore additional criteria are imposed on angular quantities in order to suppress the multijet
background: the azimuthal angle between 
$\vec{\met}$ and $\vec{\mpt}$,
$\Delta\phi(\vec{\met}, \vec{\mpt})<\pi/2$; 
the minimum azimuthal angle 
between $\vec{\met}$ and any jet, 
${\rm min}[\Delta \phi(\vec{\met},{\rm jet})] > 1.5$;
and the azimuthal angle between $\vec{\met}$ and the jet pair combination used to reconstruct the Higgs candidate,
$\Delta \phi (\vec{\met},b\bar{b})>2.8$.

It is not possible to accurately reconstruct the invariant mass of the $ZH$ system due to the missing neutrinos, so the transverse mass is used as the final
discriminant: \mbox{$m^{\text{T}}_{VH}= \sqrt{(E_{T}^{b\bar{b}} + {\met})^2 - (\ensuremath{\vec{\pt^{b\bar{b}}}}+
\vec{\met})^2}$}, where $\ensuremath{{{\pt^{b\bar{b}}}}}$ is the transverse momentum of the Higgs candidate.
The total acceptance times selection efficiency varies from 15\%  for $m_{R_1} = 400$ GeV, to 30\% for $m_{R_1}  = 1000$ GeV and down to 2\%  for $m_{R_1}  = 2000$ GeV. The drop at very high masses is due to the merging of the jets.

\subsection{ $\ell\nu b\bar{b}$ channel}

In order to suppress the multijet background and ensure the single-lepton triggers are
 fully efficient, tighter identification criteria  are placed on the lepton in this channel. 
The lepton $\pt$ requirement is raised to $\pt>25$~GeV and, for the muon channel,  the pseudorapidity  is restricted to
$|\eta|<2.5$.
Moreover, the tracking isolation is tightened and required to be less than 4\% of the lepton $\pT$.  Similarly, the sum of transverse energy deposits in the calorimeter within a cone of $\Delta R=0.3$ around the lepton, excluding the transverse energy due to the lepton and the correction for the expected pile-up contribution, is required to be less than 4\% of the lepton $\pT$.

 The multijet background is further reduced by requiring
$\Delta\phi(\vec{\met},{\rm jet})>1.0$. 
$W$ boson candidates are selected by requiring $\met>30$~GeV 
and the transverse mass  reconstructed from the
lepton and $\met$, $m^{\text T}_W = \sqrt{2 \times E_{\text T}^{\ell} \times \met \times (1-\text{cos}\Delta \phi (\ell, \vec{\met})}) >20$~GeV.

The $WH$ system mass, $m_{VH}$, is reconstructed from the lepton, the
$\met$ and the two jets. The momentum of the neutrino in the $z$-direction, $p_{z}$, is obtained by imposing the $W$ boson mass constraint  on the lepton and neutrino system, which leads to a quadratic equation. Here $p_z$ is taken as either the real component of the complex solutions or the smaller of the two real solutions.

In order to reduce the $W+$jets background, a requirement is imposed on the
transverse momentum of the $W$ boson, $p_{\text{T}}^{W}>0.4 \times m_{VH} $. The cut depends on $m_{VH}$ since the
background is generally produced at low $p_{\text{T}}^{W}$, 
whereas for signal the mean
$p_{\text{T}}^W$ increases with $m_{VH}$.
The total acceptance times selection efficiency varies from 8\% for $m_{R_1} = 400$ GeV, to 20\% for $m_{R_1}  = 1000$ GeV and down to 2\%  for $m_{R_1}  = 2000$ GeV.

\subsection{ $\ell\ell b\bar{b}$ channel}
Events in this channel are selected by requiring two reconstructed
leptons of the same flavour with opposite charge.
In order to  reduce the multijet background while keeping a high signal
acceptance, tighter requirements are placed on one
of the leptons.
These tighter electrons or muons must have $\pt>25$~GeV and, 
in addition, muons are restricted to $|\eta|<2.5$.
A cut on the two-lepton invariant mass of  
$83$~GeV$<m_{\ell\ell}<99$~GeV is imposed to reduce $t\bar{t}$ and multijet backgrounds.
The $t\bar{t}$ background is further reduced by requiring $\met<60$~GeV.

The invariant mass of the two leptons and two  jets is used
to reconstruct $m_{VH}$.

In order to reduce the dominant $Z+$jets background, a selection, optimized for this channel,
is imposed on the transverse momentum of the $Z$ boson: 
$p_{\text{T}}^{Z}>0.4\times  m_{VH}-100$~GeV. 
The total acceptance times selection efficiency varies from 18\% for $m_{R_1} = 400$ GeV, to 30\% for $m_{R_1}  = 1000$ GeV and down to 1\%  for $m_{R_1}  = 2000$ GeV.

%% file: sigbkg.tex
All backgrounds except  the multijet background are estimated from simulation, with data-based corrections for the dominant $W/Z$+jets background as described in the following.
The rate and shape of the multijet (MJ) background are estimated with data-driven
methods.

The MJ background is estimated in the 0-lepton channel using an ``ABCD method'' based on two uncorrelated variables: 
 ${\rm min}[\Delta \phi(\vec{\met},{\rm jet})]$ and $\Delta\phi(\vec{\met}, \vec{\mpt})$. The data are divided into four regions such that three of the regions are dominated by background.
 The signal region (A) is defined as explained in Sect.~\ref{sec:selection}.  The MJ-dominated region C is obtained by reversing the $\Delta\phi(\vec{\met}, \vec{\mpt})$ requirement.
 An MJ template in region A is obtained using events in region C after subtracting the contribution of other backgrounds, taken from simulation.
 The template  is then normalized by a fit to  the regions with ${\rm min}[\Delta \phi(\vec{\met},{\rm jet})] < 0.4$~\cite{AtlasVHbb} (regions B and D with orthogonal $\Delta\phi(\vec{\met}, \vec{\mpt})$ requirements).
%
%
%

In the 1-lepton channel, the MJ background is determined separately for
the electron and muon sub-channels.  An MJ-background
template is obtained from an MJ-dominated region after subtracting the small  contribution from the other backgrounds.
An MJ-dominated region is obtained by loosening the lepton identification requirements and reversing the isolation criteria.
A binned fit of the  full $\met$ spectrum of the data to the sum of the MJ contribution, $W/Z$+jets and other MC contributions is then used to extract the MJ normalization. 
The templates are validated in  a control region enriched in MJ events, selected  by reversing the $\met$ requirement.

%

For the 2-lepton channel in the $ee b \bar{b}$ final state, the MJ background shape is
determined by selecting events  with reversed electron isolation criteria  and  its normalization is extracted by fitting the full data $m_{ee}$ distribution including $Z$ sidebands.  
The MJ background in the $\mu\mu b \bar{b}$ final state is found to be negligible.

The $W/Z$+jets simulated samples are split into different components according
to the true flavour of the jets, i.e. $W/Z+qq$, $W/Z+cq$, where $q$ denotes a light quark ($u$, $d$, $s$) or a gluon, and  $W/Z$ plus heavy flavour (hf). The latter  includes:  $W/Z+b\bar{b}$, $W/Z+bq$,$W/Z+bc$, $W/Z+cc$. The normalizations of $W+cq$, $Z+cq$  and $W+$hf, $Z+$hf are free parameters of the global likelihood  fit. 
The scale factors after the fit are all consistent with 1, except for the Z+hf normalization that is 15\% higher as seen in previous measurements~\cite{AtlasVHbb}.
The $W/Z$+jets modelling is checked in  control regions selected by requiring events with no $b$-tagged jets or in the  $m_{b\bar{b}}$ sideband region in the 1-tag and 2-tag channels.
A difference between data and simulation is observed in the 0-tag control region and a correction is extracted  as a function of the azimuthal angle difference between  the two leading-$\pT$ jets, $\dphi$.
 This is used to reweight the $Z+qq$ and $W+qq$ components. After this correction is applied a discrepancy is observed in the  $\pt^{\ell\ell}$ distribution in the 2-lepton channel after the requirement of at least one $b$-tagged jet. A correction is extracted and used to reweight the $Z+cq$ and $Z+$hf components. The full procedure is described in detail in Ref.~\cite{AtlasVHbb}.
  
The background contributions from single top quark and diboson production are normalized to the number of background events predicted by simulation while the $t\bar{t}$ normalization is a free parameter in the likelihood fit. 
The description of the shape of the $t \bar{t}$ background from MC simulation has been validated in  samples dominated by top pair events. Good agreement within uncertainties is observed between data and expectation in these validation regions.

 The $t \bar{t}$ control region is   defined by requiring exactly one electron and one muon, one of which has $\pt>25$~GeV,
and two $b$-tagged jets.  It is included in the likelihood fit to constrain the $t \bar{t}$ normalization. The scale factor for the $t\bar{t}$ normalization  is found to be  $1.03 \pm 0.04$ after the likelihood fit to the 0- and 2-lepton channel plus the $t \bar{t}$ control region, and $0.99 \pm 0.09$ from the fit to the 1-lepton channel.  The fit procedure is 
 described in more detail in Sect.~\ref{sec:results}.

%% file: systs.tex
The most important experimental systematic uncertainties
come from the jet energy scale (JES)  and $b$-tagging
efficiency.
	
The JES  systematic uncertainty arises from several sources including
uncertainties from the in-situ calibration, the corrections dependent on pile-up 
and the jet flavour composition~\cite{Aad:2014bia}. The fractional systematic uncertainty on the JES ranges from 3\% for a  20 GeV jet to 1\% for a 1 TeV jet.

The uncertainty due to the jet energy
resolution is also considered. It varies from 20\% for a jet with $\pt >$ 20 GeV  to 5\% for a  jet with $\pt >$ 1 TeV. The jet energy scale and resolution
uncertainties are propagated to the reconstructed $\vec{\met}$. The
uncertainty on $\vec{\met}$  also has  a contribution from hadronic energy
that is not included in jets~\cite{ATLAS-CONF-2013-082}.

The $b$-tagging efficiency uncertainty depends on jet $\pt$
and comes mainly from the uncertainty on the measurement of the
efficiency in $t\bar{t}$ events~\cite{ATLAS-CONF-2014-004}.
Uncertainties are also derived for $c$- and light-flavour jet tagging~\cite{ATLAS-CONF-2014-046}.
	
Other experimental systematic uncertainties that have a smaller impact are those on the lepton energy scale and identification efficiency and the
efficiency of the triggers.

In addition to the experimental systematic uncertainties, uncertainties are taken into account for possible differences between  data and
the simulation model that is used for each process.  For the background modelling uncertainties the procedure described
in Ref.~\cite{AtlasVHbb} is followed. The $Z$+jets and $W$+jets backgrounds include
uncertainties on the relative fraction of the different flavour components, and shape uncertainties on the modelling of $m_{b\bar{b}}$, $\dphi$ and $\pt^{Z}$ distributions.
For $t\bar{t}$ production, shape uncertainties are included for the modelling of  top quark transverse momentum, $m_{b\bar{b}}$  and $m_{VH}$ distributions.
The uncertainty on the MJ background shape in the 1-lepton channel is evaluated by using alternative templates obtained by changing the definition of the data sidebands.
The uncertainty on the MJ background normalization is taken to be 100\%, 30\% and 50\% for the 0-, 1- and 2-lepton  channels, respectively. These are extracted from fits using alternative templates.	

The dominant uncertainties on the signal acceptance arise from the choice of PDFs (2--5\%) estimated by comparing the default PDFs to other sets, and from the factorization and renormalization scales (5--10\%) obtained by varying these up and down by a factor of two.


%% file: results.tex
\begin{figure*}[t!]
\centering
\subfloat[$\nu\nu b \bar{b}$]{
  \includegraphics[width=0.31\textwidth]{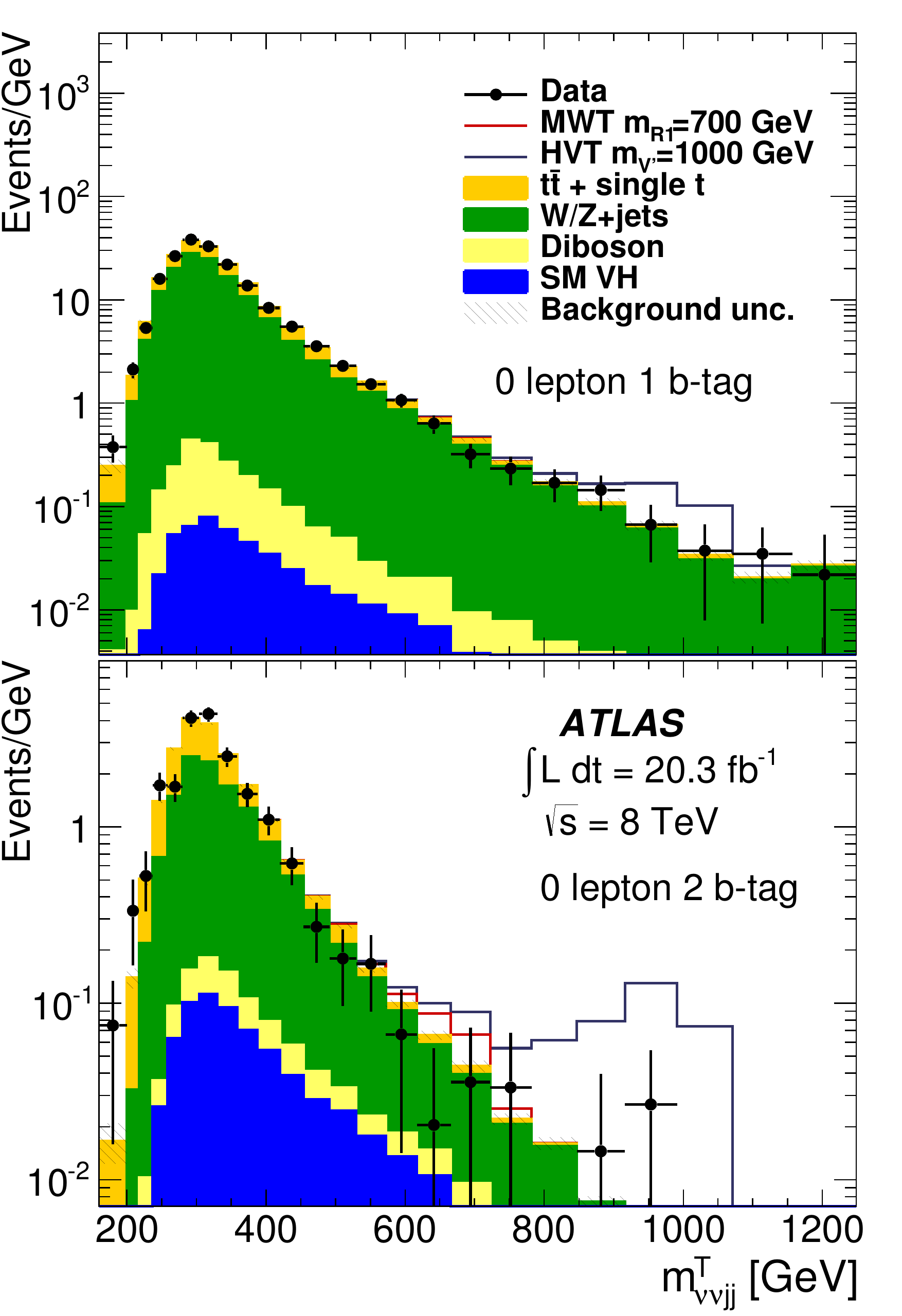}
}
\subfloat[$\ell \nu b \bar{b}$]{
  \includegraphics[width=0.31\textwidth]{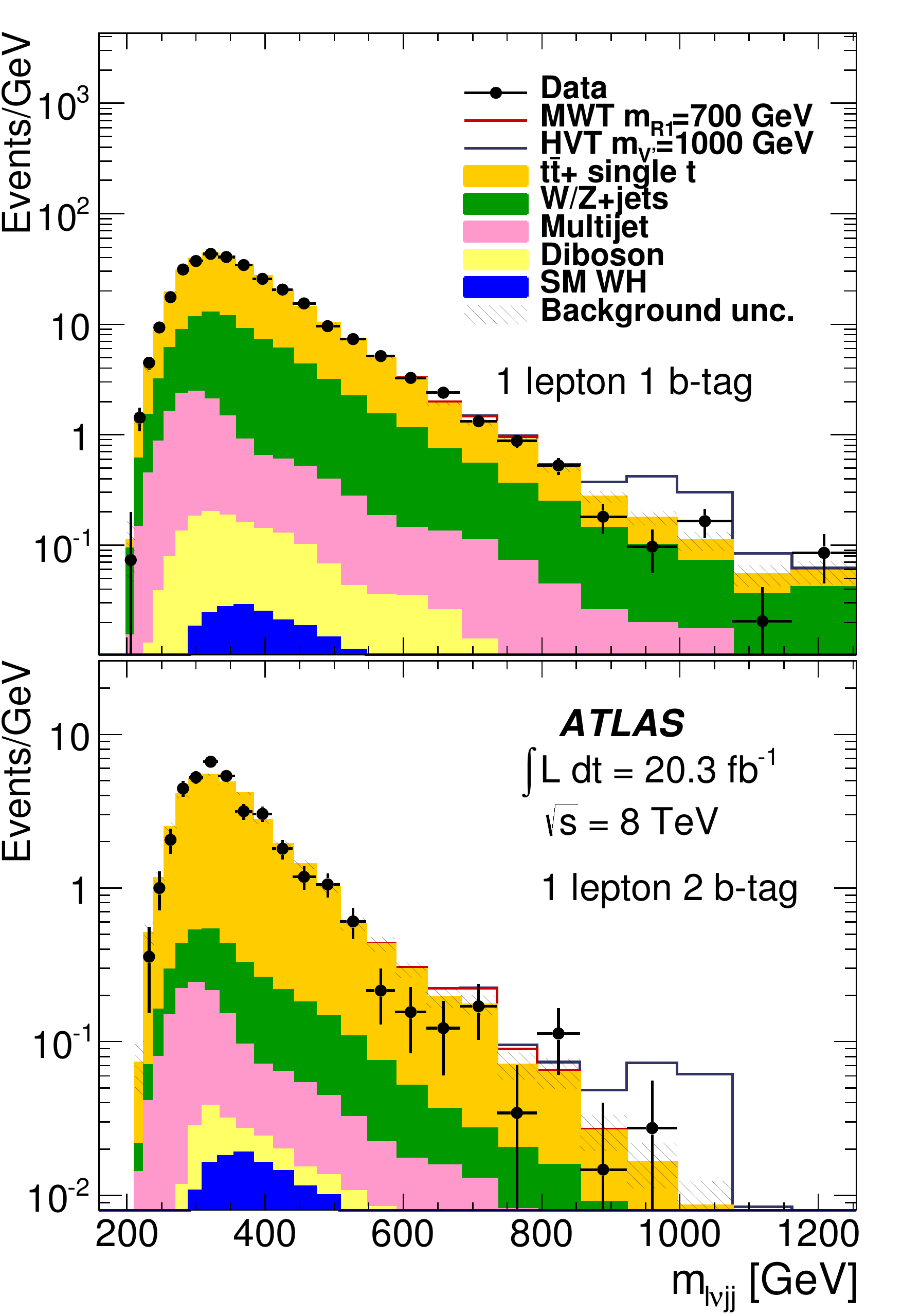}
}
\subfloat[$\ell\ell b \bar{b}$]{
  \includegraphics[width=0.31\textwidth]{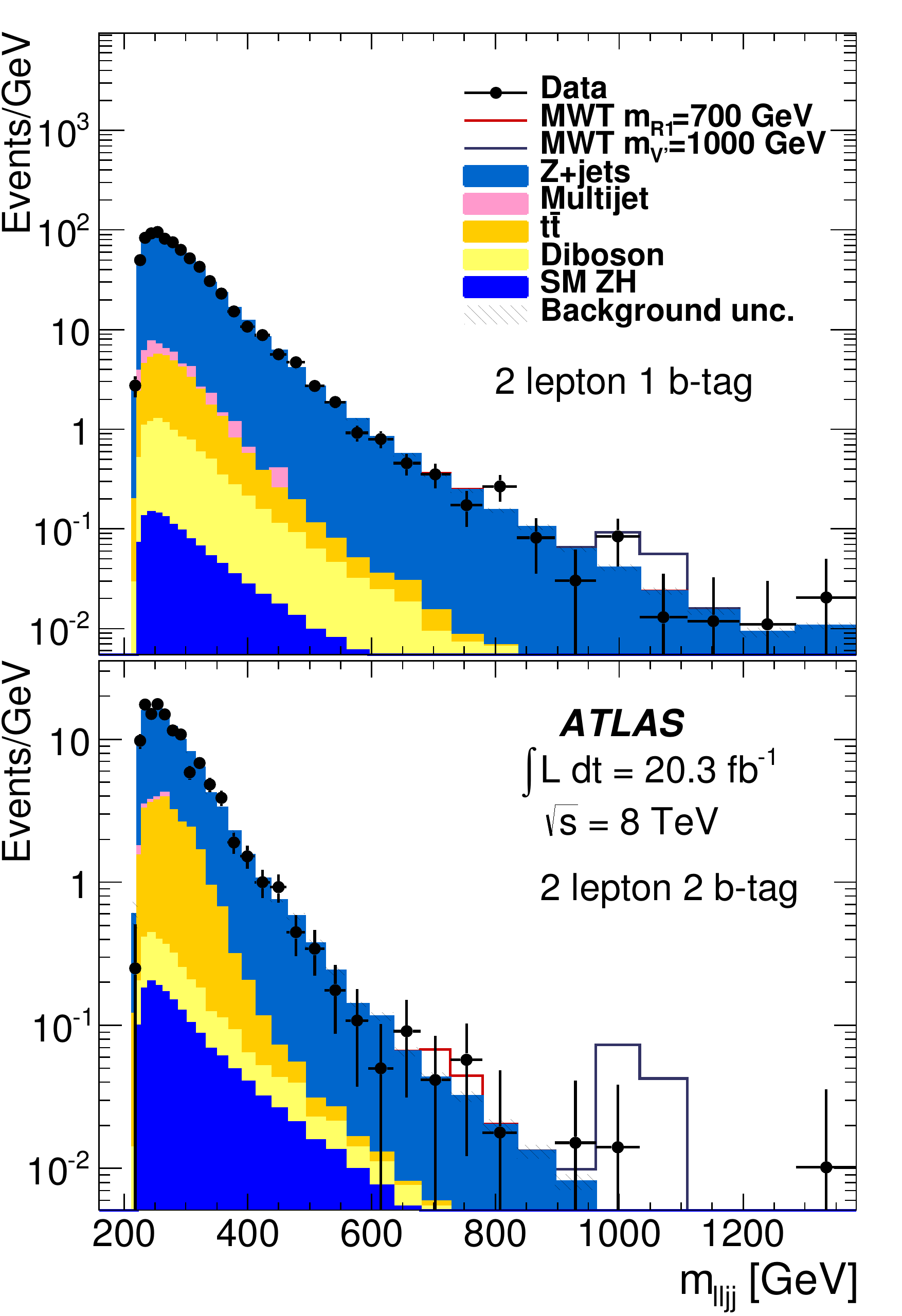}
}

  \caption{Distributions of the reconstructed  (a) transverse mass $m^{\text{T}}_{\nu\nu jj}$ for the $\nu\nu b \bar{b}$ final state, (b) invariant mass $m_{\ell \nu jj}$ for the $\ell \nu b \bar{b}$  final state	and  (c)  invariant mass $m_{\ell \ell jj}$ for the $\ell\ell b \bar{b}$  final state for the 1-$b$-tag (upper) and  2-$b$-tag (lower) channels. 
	The background expectation is shown after the profile likelihood fit to the data. 
  Any overflow is included in the last bin.	
	The signals are shown stacked on top of the background and correspond to the benchmark models MWT with $m_{R1} = 700$~GeV and HVT with $m_{V'} = 1000$~GeV normalized to the expected cross sections.
	}
	  \label{fig:xxbb_final}
	  	\end{figure*}
\begin{table}
   \caption{The number of expected and observed
 events for the three final states. The expectation  is shown 
 after the profile likelihood fit to the data. The quoted uncertainties are the combined
systematic and statistical uncertainties. The overall background is more constrained than the individual components, causing the errors of individual components to be anti-correlated.}
   \centering
   \begin{footnotesize}
   \begin{tabular}{lccc}
\hline
	\multicolumn{4}{c}{Two $b$-tags}\\
	\hline
	           &    $\nu\nu b \bar{b}$ & $\ell \nu b \bar{b}$ & $\ell\ell  b \bar{b}$            \\
	 \hline
	   $Z$+jets    & \phantom{.}224  $\pm$ 14\phantom{.}       &  \phantom{0}3.2 $\pm$ 0.2 & \phantom{.}1198  $\pm$ 47\phantom{.}  \\
	 $W$+jets & \phantom{0.}82  $\pm$  29\phantom{.}     &   \phantom{0.}61 $\pm$    21\phantom{.}  &     \phantom{00}--        \\ 
	 $\ttbar$ & \phantom{.}166 $\pm$ 10\phantom{.}      &  \phantom{.}718 $\pm$ 42\phantom{.}& \phantom{0.}321 $\pm$ 14\phantom{.} \\ 
  Single top & 23.2  $\pm$   2.6 &  71.3 $\pm$ 8.1&       \phantom{00}--    \\ 
	 Diboson & 10.1  $\pm$ 1.1   &    \phantom{0}2.8 $\pm$ 0.6& \phantom{0}25.9 $\pm$ 5.8    \\ 
	SM $VH$    & 20.3 $\pm$ 8.1     &   \phantom{0}4.6 $\pm$ 1.9  & \phantom{0}24.4 $\pm$ 6.1    \\ 
    Multijet & $<$ 3     &  \phantom{0.}29 $\pm$ 13\phantom{.}    & \phantom{0}12.1 $\pm$ 9.1\\ 
\hline	
	Total bkg. & \phantom{.}524  $\pm$ 20\phantom{.}    & \phantom{.}889  $\pm$ 28\phantom{.}  & \phantom{.}1581  $\pm$ 39\phantom{.}  \\
	\hline
		Data  &   511     &  879   &     1593     \\
	\hline
	\hline
	\multicolumn{4}{c}{One $b$-tag}\\
  	\hline
	$Z$+jets  & 2120 $\pm$ 150\phantom{.}       &  53.6 $\pm$  4.7\phantom{.}                          & 9120  $\pm$ 550\\
	$W$+jets & 1450  $\pm$  360\phantom{.}     &   1892 $\pm$ 590\phantom{.}  &     --        \\ 
	$\ttbar$   & 928 $\pm$ 43\phantom{.}           &  4650 $\pm$ 370\phantom{.}   & 425 $\pm$ 10 \\ 
    Single top & 221 $\pm$ 25\phantom{.}           &  832 $\pm$ 94\phantom{.}                          &      --    \\ 
	Diboson   & 48.2 $\pm$ 3.8\phantom{.}         &   32.1 $\pm$ 2.5\phantom{.}    & 124 $\pm$ 17    \\ 
	SM $VH$      & 13.4 $\pm$ 5.2\phantom{.}         &  \phantom{0}6.9 $\pm$ 2.7\phantom{.}      & 17.6 $\pm$ 7.0  \\ 
	Multijet   & $<$ 10                                &  316 $\pm$ 83\phantom{.}     & 139 $\pm$ 70\\ 
\hline	
	Total bkg. & 4761 $\pm$ 69\phantom{..}    & 7782 $\pm$ 97\phantom{..}  & 9830  $\pm$ 100  \\
	\hline
		Data  &   4758     &  7854  &     9827     \\
	\hline

\end{tabular}	   
	\end{footnotesize}
	   	   \label{tab:yield}
	 \end{table}

\begin{figure*}[ht!]
	  \centering
	
	  	 \subfloat[$R_1^{0}  (V'^{0})  \rightarrow ZH$, $H \rightarrow b\bar{b}$]{
	    \includegraphics[width=0.45\textwidth]{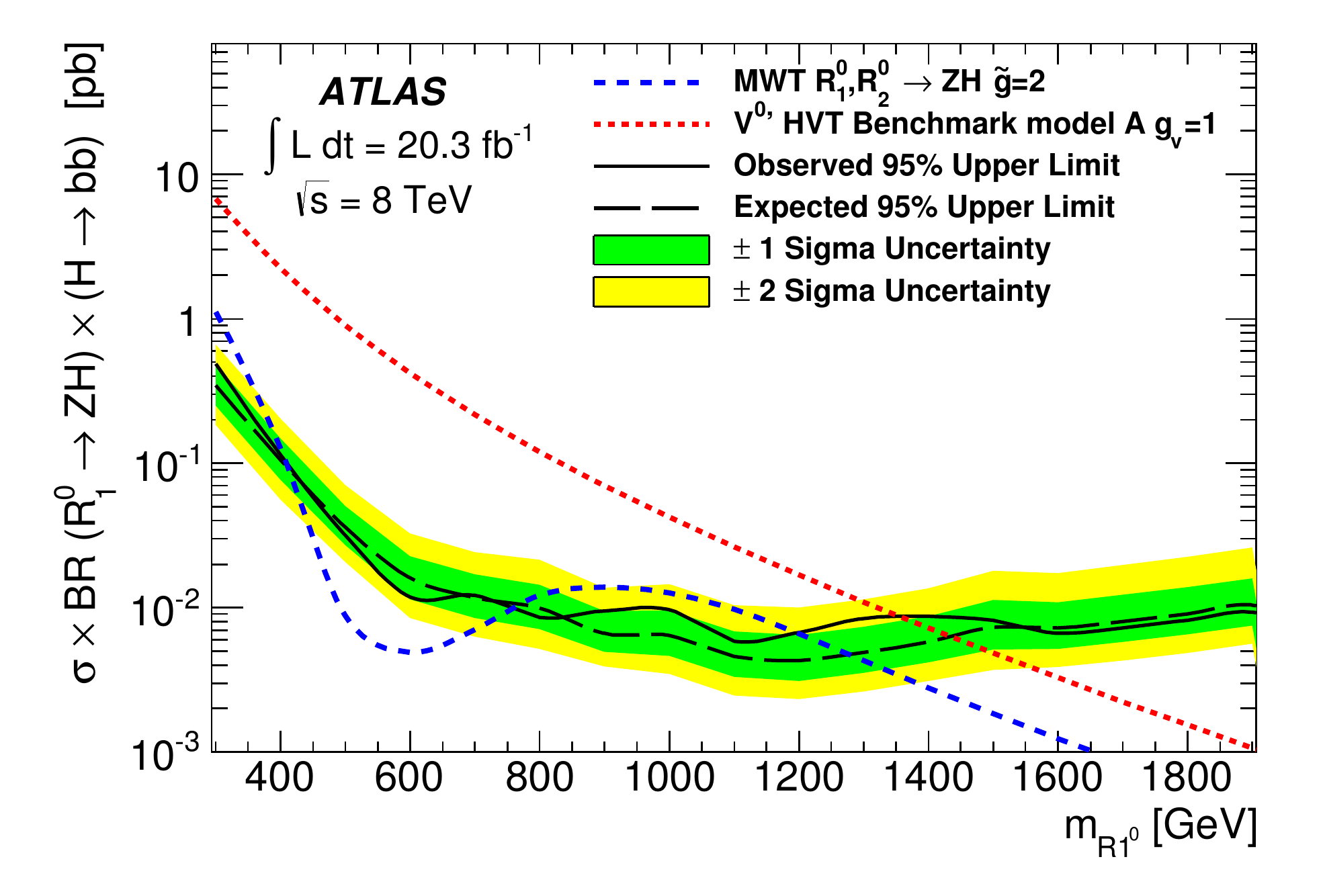}
	  }
	 \subfloat[$R_1^{\pm} (V'^{\pm})  \rightarrow WH$, $H \rightarrow b\bar{b}$]{
	    \includegraphics[width=0.45\textwidth]{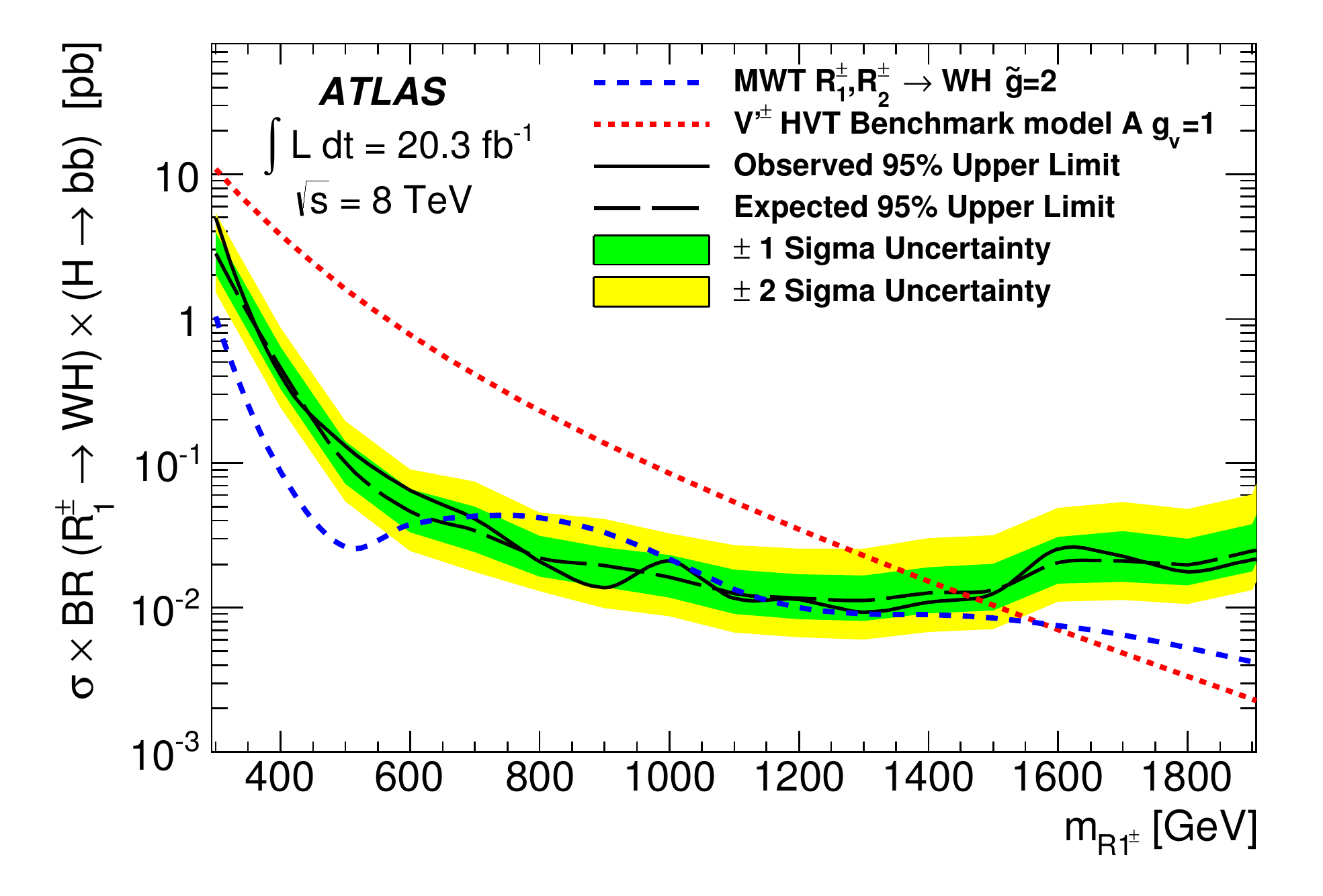}
	  }
	  \caption{Combined upper limits at the 95\% CL for (a)  the 	production
	cross section of $R_1^{0}$ ($V'^{0}$)   times its branching ratio to $ZH$ and branching ratio of  $H$ to $b\bar{b}$ and (b) the production
	cross section of $R_1^{\pm}$  ($V'^{\pm}$)   times its branching ratio to $WH$ and branching ratio of  $H$ to $b\bar{b}$ . The experimental limits are obtained using samples with a single resonance $R_1$; however, the theory curve line for MWT includes both $R_1$ and $R_2$. The dip  near 500 GeV in this theory curve is due to the interference between $R_1$ and $R_2$~\cite{Belyaev:2008yj}. }
	  \label{fig:limit}
	\end{figure*} 

The reconstructed mass distributions for events passing the selection are shown in Fig.~\ref{fig:xxbb_final}.  The background  expectation is shown after the profile likelihood fit to the data. Table~\ref{tab:yield} shows the number of events expected and observed in each final state.

No significant excess of events is observed in the data compared to the prediction from SM background sources.
Exclusion limits at the 95\% confidence level (CL) are set on the production cross section times the branching fraction  for MWT and HVT models.
The limits for the charged resonance  are obtained by performing the likelihood fit over the $\ell\nu b\bar{b}$ channel alone, while the $\ell\ell b\bar{b}$, $\nu\nu b\bar{b}$ channels  as well as the  $t\bar{t}$ control region are used for the neutral resonance.

\begin{figure}
  \centering	
	    \includegraphics[width=0.45\textwidth]{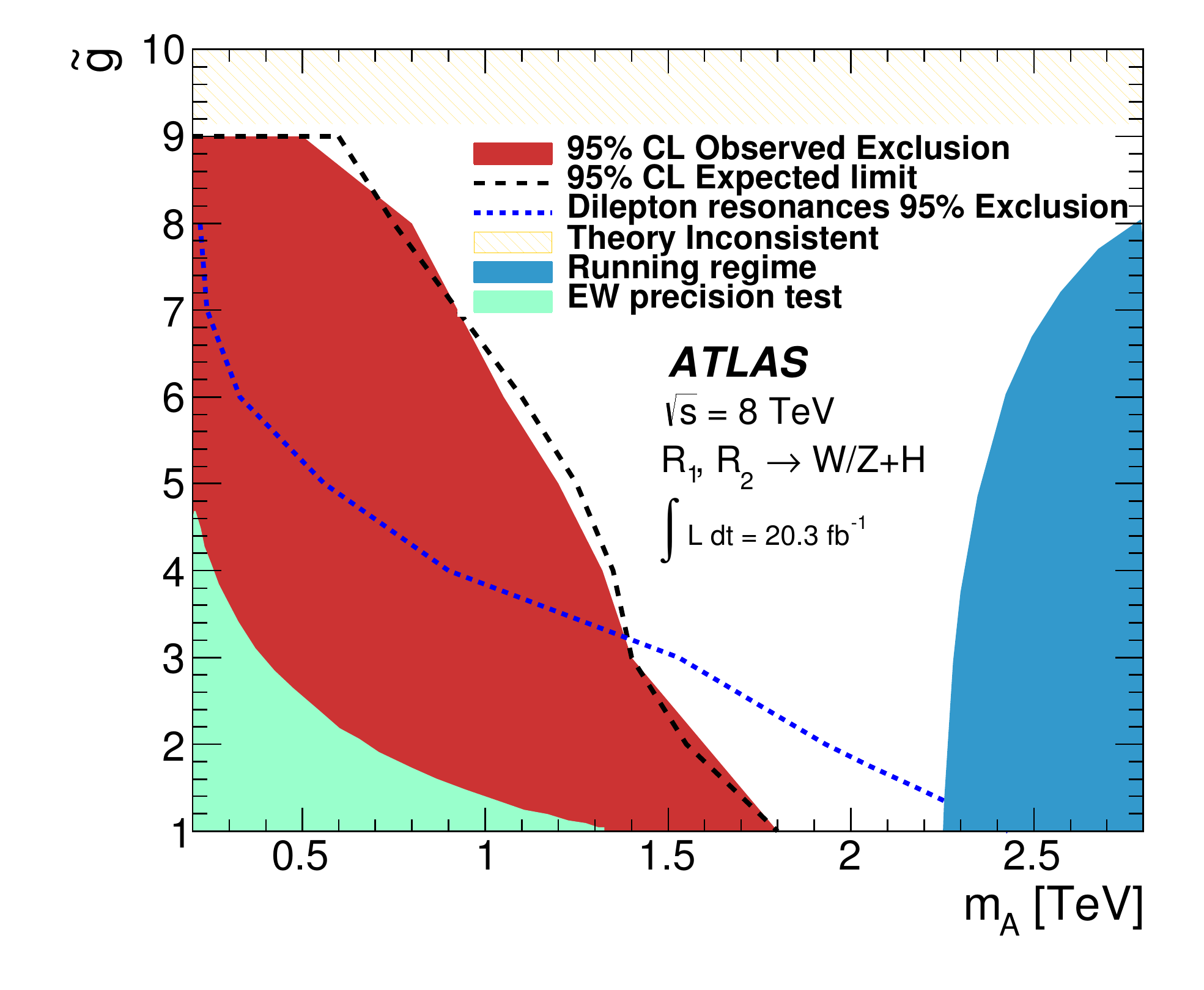}	  
	  \caption{Exclusion contours at 95\% CL in the plane of
the Minimal Walking Technicolor parameter space defined by
the bare axial-vector mass versus the strength of the spin-1
resonance interaction \{$m_A$, $\tilde{g}$\}. Electroweak precision measurements
exclude the (green) area in the bottom left corner.
The requirement to stay in the walking regime excludes the
(blue) area in the right corner. The large (red) area (black dashed
line) shows the observed (expected) exclusion. The blue dashed line
shows the observed exclusion from the dilepton resonance
search~\cite{Aad:2014cka}. The upper region is excluded due to non-real
axial and axial-vector decay constants. Here both resonances predicted by MWT, $R_{1}$ and $R_{2}$, are fitted simultaneously.}
	  \label{fig:Mavsgtilde}
	\end{figure}

The exclusion limits are calculated with a modified frequentist method \cite{CLs_2002}, also known as \textit{CL$_{s}$},
and the profile-likelihood test statistic \cite{Cowan2011}, using the binned $m_{\text{VH}}$  mass distributions for $\ell\nu b\bar{b}$, $\ell\ell b\bar{b}$ and  $\nu\nu b\bar{b}$ final states.
Systematic uncertainties and their correlations are taken into account as nuisance parameters. None of the systematic uncertainties considered are significantly constrained or pulled in the likelihood fit.
Figure~\ref{fig:limit} shows 95\% CL upper limits on the production
cross section multiplied by the branching fraction into $WH$ and $ZH$
as a function of the resonance mass separately for the charged
$R_1^{\pm}$ and for the neutral $R_1^{0}$.  The experimental limits
are obtained using samples with a single resonance $R_1$,  where the cross section for $R_2$ has been set to zero to be less model-dependent. 
The theoretical predictions for the HVT benchmark \textit{model A} with coupling constant $g_{V}=1$
allow exclusion of $m_{V'^{0}} <1360$~GeV ($m_{V'^{\pm}} <1470$~GeV). For the MWT model, since there are two resonances of different mass, the results are displayed for the first one, $R_1^{0,\pm}$.  The excluded regions are  $m_{R_1^{0}}<410$~GeV, $750<m_{R_1^{0}}<1200$~GeV ($700<m_{R_1^{\pm}}<1150$~GeV).
The dip near $500$ GeV in this theory curve is due to the interference between $R_1$ and $R_2$~\cite{Belyaev:2008yj}.  
To study the scenario in which  the masses of charged and neutral resonances are the same,  a combined likelihood fit over all signal regions and the $t\bar{t}$ control region is also performed. The exclusion contours in the \{$m_{A}$,$\tilde{g}$\} plane for MWT are presented in Fig.~\ref{fig:Mavsgtilde}. For this result,  both resonances predicted by MWT, $R_1$ and $R_2$, are fitted simultaneously and, at each $\tilde{g}$, the different branching ratios to $WH$ and $ZH$ are taken into account. Electroweak precision data, a requirement to remain within the walking technicolor regime and constraints from requiring real-valued physical decay constants exclude a portion of the plane.  This analysis is particularly sensitive at high $\tilde{g}$ values, where the limits exceed those from the dilepton resonance search~\cite{Aad:2014cka}.

\begin{figure}
  \centering	
	    \includegraphics[width=0.48\textwidth]{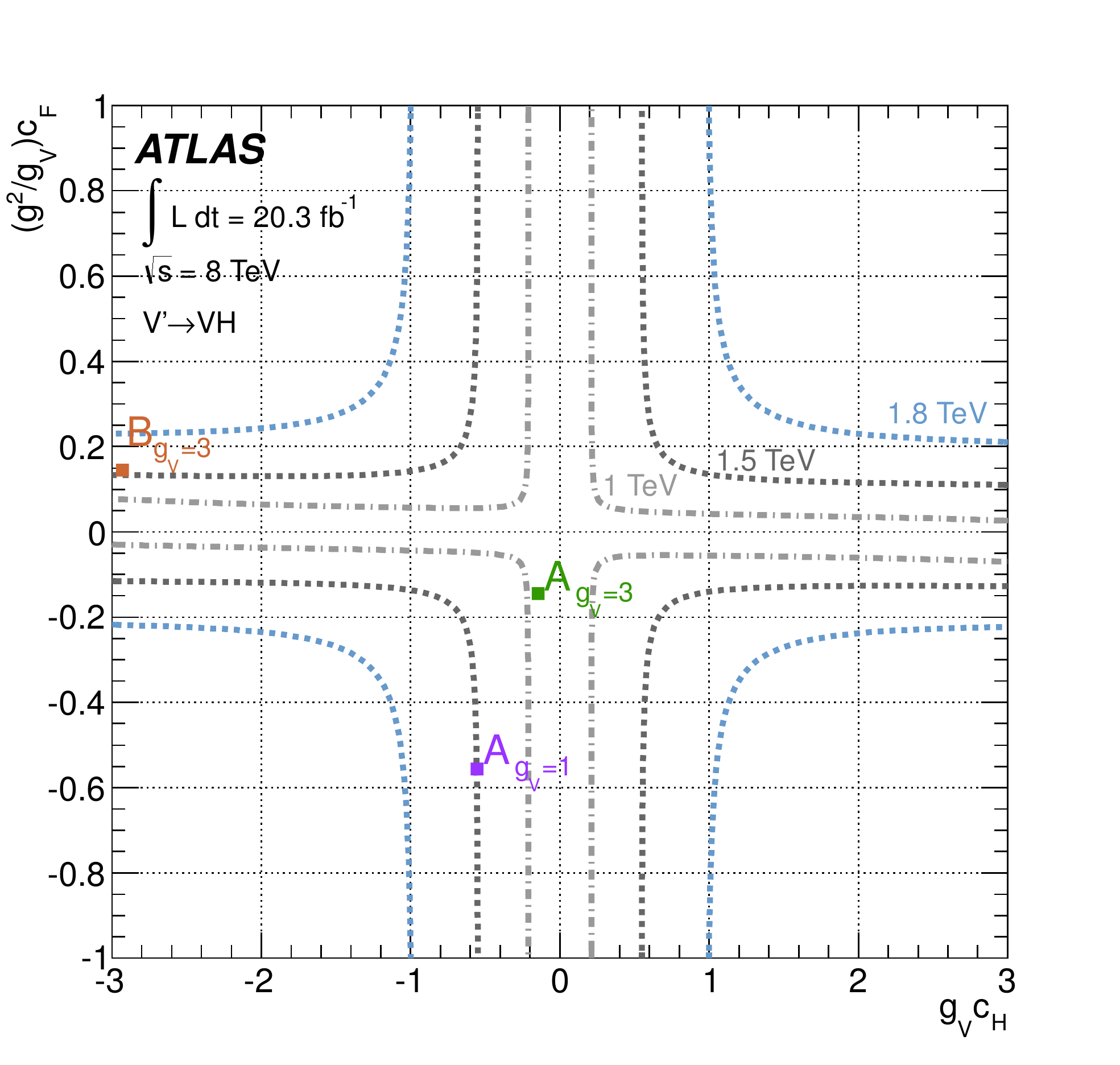}	    
	  \caption{Observed 95\% CL exclusion contours in the HVT parameter space \{($g^2$/$g_V$)$c_F$, $g_Vc_H$\} for resonances of mass 1~\TeV, 1.5~\TeV\ and 1.8~\TeV\ . The areas outside the curves are excluded.  Also shown are the benchmark model parameters A($g_V$=1), A($g_V$=3) and B($g_V$=3).}
	  \label{fig:HVT_contours}
	\end{figure}

The exclusion contours  in the HVT parameter space \{($g^2$/$g_V$)$c_F$, $g_Vc_H$\} for resonances of mass 1~\TeV, 1.5~\TeV\ and 1.8~\TeV\ are shown in Fig.~\ref{fig:HVT_contours} where all three channels are combined, taking into account the branching ratios to $WH$ and $ZH$ from the HVT model. These contours are produced by scanning the parameter space, using the HVT tools provided in a web-interface~\cite{HVT-web,Pappadopulo:2014qza}.

%% file: summary.tex
A search for a new heavy resonance decaying to $WH/ZH$ is presented in this Letter.
The search is performed using  20.3 fb$^{-1}$ of $pp$ collision data at  8 TeV centre-of-mass energy collected by the ATLAS detector at the Large Hadron Collider.
No significant deviations from the SM background predictions are observed in the three final states considered:  $\ell \ell b \bar{b}$, $\ell \nu b \bar{b}$, $\nu \nu b \bar{b}$.
Upper limits are set at the 95\% confidence level on the production cross sections of $R_1$ and $V'$ for the Minimal Walking Technicolor and Heavy Vector Triplets models respectively.  Exclusion contours at 95\% CL in the MWT parameter space \{$m_{A}$,$\tilde{g}$\} and  in the HVT parameter space \{($g^2$/$g_V$)$c_F$, $g_Vc_H$\} are  presented.

%% file: acknowledgements/Acknowledgements.tex

We thank CERN for the very successful operation of the LHC, as well as the
support staff from our institutions without whom ATLAS could not be
operated efficiently.

We acknowledge the support of ANPCyT, Argentina; YerPhI, Armenia; ARC,
Australia; BMWFW and FWF, Austria; ANAS, Azerbaijan; SSTC, Belarus; CNPq and FAPESP,
Brazil; NSERC, NRC and CFI, Canada; CERN; CONICYT, Chile; CAS, MOST and NSFC,
China; COLCIENCIAS, Colombia; MSMT CR, MPO CR and VSC CR, Czech Republic;
DNRF, DNSRC and Lundbeck Foundation, Denmark; EPLANET, ERC and NSRF, European Union;
IN2P3-CNRS, CEA-DSM/IRFU, France; GNSF, Georgia; BMBF, DFG, HGF, MPG and AvH
Foundation, Germany; GSRT and NSRF, Greece; RGC, Hong Kong SAR, China; ISF, MINERVA, GIF, I-CORE and Benoziyo Center, Israel; INFN, Italy; MEXT and JSPS, Japan; CNRST, Morocco; FOM and NWO, Netherlands; BRF and RCN, Norway; MNiSW and NCN, Poland; GRICES and FCT, Portugal; MNE/IFA, Romania; MES of Russia and NRC KI, Russian Federation; JINR; MSTD,
Serbia; MSSR, Slovakia; ARRS and MIZ\v{S}, Slovenia; DST/NRF, South Africa;
MINECO, Spain; SRC and Wallenberg Foundation, Sweden; SER, SNSF and Cantons of
Bern and Geneva, Switzerland; NSC, Taiwan; TAEK, Turkey; STFC, the Royal
Society and Leverhulme Trust, United Kingdom; DOE and NSF, United States of
America.

The crucial computing support from all WLCG partners is acknowledged
gratefully, in particular from CERN and the ATLAS Tier-1 facilities at
TRIUMF (Canada), NDGF (Denmark, Norway, Sweden), CC-IN2P3 (France),
KIT/GridKA (Germany), INFN-CNAF (Italy), NL-T1 (Netherlands), PIC (Spain),
ASGC (Taiwan), RAL (UK) and BNL (USA) and in the Tier-2 facilities
worldwide.

%% file: atlas_authlist.tex
\begin{flushleft}
{\Large The ATLAS Collaboration}

\bigskip

G.~Aad$^{\rm 85}$,
B.~Abbott$^{\rm 113}$,
J.~Abdallah$^{\rm 151}$,
O.~Abdinov$^{\rm 11}$,
R.~Aben$^{\rm 107}$,
M.~Abolins$^{\rm 90}$,
O.S.~AbouZeid$^{\rm 158}$,
H.~Abramowicz$^{\rm 153}$,
H.~Abreu$^{\rm 152}$,
R.~Abreu$^{\rm 30}$
% ...
\bigskip
\\
% ...
$^{11}$ Institute of Physics, Azerbaijan Academy of Sciences, Baku, Azerbaijan\\
% ...
$^{30}$ CERN, Geneva, Switzerland\\
% ...
$^{85}$ CPPM, Aix-Marseille Universit{\'e} and CNRS/IN2P3, Marseille, France\\
% ...
$^{90}$ Department of Physics and Astronomy, Michigan State University, East Lansing MI, United States of America\\
% ...
$^{107}$ Nikhef National Institute for Subatomic Physics and University of Amsterdam, Amsterdam, Netherlands\\
% ...
$^{113}$ Homer L. Dodge Department of Physics and Astronomy, University of Oklahoma, Norman OK, United States of America\\
% ...
$^{151}$ Institute of Physics, Academia Sinica, Taipei, Taiwan\\
$^{152}$ Department of Physics, Technion: Israel Institute of Technology, Haifa, Israel\\
$^{153}$ Raymond and Beverly Sackler School of Physics and Astronomy, Tel Aviv University, Tel Aviv, Israel\\
% ...
$^{158}$ Department of Physics, University of Toronto, Toronto ON, Canada\\
% ...
\end{flushleft}
